\documentclass[reprint, twocolumn, amsmath, amssymb, showpacs, superscriptaddress, aps, prl]{revtex4-2}

\usepackage{microtype}
\usepackage{graphicx}
\usepackage{dcolumn}
\usepackage{bm}
\usepackage[colorlinks=true,
						linkcolor=blue,
						urlcolor=blue,
						citecolor=blue,
						bookmarks=true,
						pdfborder={0 0 0}]{hyperref}
\usepackage{multirow}
\usepackage{xcolor}
\usepackage[export]{adjustbox}
\usepackage{tikz}
\usetikzlibrary{shapes.geometric}

\definecolor{mypurple}{RGB}{136, 0, 136}
\definecolor{myblue}{RGB}{27, 189, 255}
\definecolor{mygreen}{RGB}{54, 221, 89}
\definecolor{mypink}{RGB}{255, 145, 175}

\newcommand{\tikzpoly}[3]{
\begin{tikzpicture}
    \node[regular polygon,
    draw = #2, fill = #2,
    regular polygon sides = #1, minimum size = 2pt, scale = #3] (0,0) {};
\end{tikzpicture} }

\newcommand{\tikzdiamond}[2]{
\begin{tikzpicture}
    \node[diamond, draw = #1, fill = #1, minimum size = 2pt, scale = #2] (0,0) {};
\end{tikzpicture} }

\newcommand{\tikzcircle}[2][red,fill=red]{\tikz[baseline=-0.5ex]\draw[#1,radius=#2] (0,0) circle ;}

\newcommand{\Gpayoff}{\tikzcircle[mypurple, fill=mypurple]{2.3pt}}
\newcommand{\Wpayoff}{\tikzdiamond{mypink}{0.4}}
\newcommand{\Spayoff}{\tikzpoly{4}{mygreen}{0.44}}
\newcommand{\Tpayoff}{\tikzpoly{3}{myblue}{0.3}}

\graphicspath{{figures/}}

\begin{document}
\title{Explosive cooperation in  social dilemmas on higher-order networks}

\author{Andrea Civilini}
\affiliation{School of Mathematical Sciences, Queen Mary University of London, London E1 4NS, United Kingdom}

\author{Onkar Sadekar}
\affiliation{Central European University Vienna, Vienna 1100, Austria}

\author{Federico Battiston}
\affiliation{Central European University Vienna, Vienna 1100, Austria}

\author{Jesús Gómez-Gardeñes}
\affiliation{Department of Condensed Matter Physics, University of Zaragoza, 50009 Zaragoza, Spain}
\affiliation{GOTHAM lab, Institute of Biocomputation and Physics of
Complex Systems (BIFI), University of Zaragoza, 50018 Zaragoza, Spain}

\author{Vito Latora}
\affiliation{School of Mathematical Sciences, Queen Mary University of London, London E1 4NS, United Kingdom}
\affiliation{Dipartimento di Fisica ed Astronomia, Universit\`a di Catania and INFN, Catania I-95123, Italy}
\affiliation{Complexity Science Hub Vienna, A-1080 Vienna, Austria}

\begin{abstract}

Understanding how cooperative behaviours can emerge from competitive interactions is an open problem in biology and social sciences. While  interactions are usually modelled as pairwise networks, the units of many real-world systems can also interact in groups of three or more. Here, we introduce a general framework to extend pairwise games to higher-order networks. 
By studying social dilemmas
on hypergraphs with a tunable structure, we find an explosive transition to cooperation triggered by a critical number of higher-order games. The associated bistable regime implies that an initial critical mass of cooperators is also required for the emergence of prosocial behavior. 
Our results show that higher-order interactions provide a novel explanation for the survival of cooperation.

\end{abstract}

\maketitle

\paragraph*{Introduction.}
The pervasiveness of cooperation in our world has long puzzled researchers \cite{axelrod_evolution_1981, nowak_supercooperators_2012}. After all, the natural world (and human society is not an exception) obeys Darwinian selection,  which is driven by the self-interest of individuals. In such a competitive world, costly altruistic cooperative behaviours seem inappropriate, since they do not bring any immediate advantage to the cooperators in the brutal fight for the survival of the fittest \cite{smith_logic_1973, nowak_evolutionary_1992, weibull_evolutionary_2004, nowak_evolutionary_2006}.
It is instead more profitable for a self-interested individual to defect (i.e. not participating to the costly altruistic behaviours), taking advantage of the benefits from the actions of cooperators who, in turn, see their sustainability jeopardized by the higher profits of free-riders \cite{hofbauer_evolutionary_1998, perc_statistical_2017}.

A well-known theoretical framework for studying the problem of the survival of cooperative traits in human societies is that of social dilemmas or collective action problems in which, given a set of actors, each of them can choose between two strategies, either to cooperate or to defect \cite{szabo_evolutionary_1998, doebeli_evolutionary_2004}. In this context, a defector receives a higher payoff than a cooperator when the two interact, but if everyone adopts the more profitable selfish strategy of defection, the payoff of the agents vanishes \cite{hardin_tragedy_1968, axelrod_further_1988, milinski_reputation_2002, dawkins_selfish_2006}. 
Social dilemma scenarios are typically studied in evolutionary game theory \cite{smith_logic_1973, smith_1982, hofbauer_evolutionary_1998, Gintins_2000, traulsen_EGTcooperation_2006, evogamefinitepop} by implementing games, such as the Prisoner's Dilemma (PD), on structured populations \cite{szabo_evolutionary_2007, allen_evolutionary_2017, antonioni_coevolution_2017}. 
The underlying structure of a population is usually modeled as a network, where links represent the interactions between pairs of agents \cite{boccaletti_complex_2006, newman_networks_2010, latora_complex_2017}.  
In some cases, the structure of the network has been shown to promote prosocial behaviours through, e.g., mechanisms of  network reciprocity \cite{nowak_evolutionary_1992, lieberman_evolutionary_2005, nowak_five_2006}, the heterogeneity of the nodes \cite{ santos_scale-free_2005, santos_evolutionary_2006, gomez-gardenes_dynamical_2007} and the presence of clustering \cite{assenza_2008_clustering}.
Despite their contributions to our comprehension of social dilemmas, these attempts to consider realistic interaction structures are limited in their representation of real-world systems.
The links of a network can indeed only describe pairwise interactions, while the units of a complex system can also interact in groups of more than two. Thus, networks do not allow to accommodate more realistic and general forms of higher-order social interactions.

In the last years, different higher-order mathematical structures, such as hypergraphs and simplicial complexes, have started to be used to represent interactions among three or more units \cite{battiston_networks_2020, battiston_physics_2021, stramaglia_disentangling_2022}. From contagion processes \cite{iacopini_simplicial_2019} to synchronization dynamics \cite{gambuzza_stability_2021, stramaglia_quantifying_2021, gallo_synchronization_2022} and ecological competition \cite{grilli_higher-order_2017}, many studies have illustrated that higher-order interactions can give rise to the emergence of novel collective behaviors and dynamical patterns not observable in pairwise networks.
The first steps have been moved to consider higher-order interactions also in the context of evolutionary games. However, all the works dealing with  $n$-person social dilemmas \cite{perc_evolutionary_2013} either still rely on pairwise networks to define group interactions, without the flexibility and generality of real higher-order networks
\cite{santos_social_2008}, or make too strong assumptions (e.g. regarding the payoff structure) that can be justified only in the particular scenario under investigation \cite{civilini_evolutionary_2021, guo_evolutionary_2021, xu_multi-player_2022, gomez-gardenes_PGG, alvarez-rodriguez_evolutionary_2021}.
For example, when hyperedges have been used to describe group interactions at the microscale level  \cite{gomez-gardenes_PGG, alvarez-rodriguez_evolutionary_2021}, the payoff associated to each group is a linear function of the strategies of the group members, hence seriously limiting the general representation of the dynamics of social dilemmas. 
Conversely, when more general payoff structures have been adopted in an extended framework of $n$-person games, only unstructured (well-mixed) populations composed of groups of the same size have been considered, and with the main focus on games with more than 2 strategies, thus not addressing the case of social dilemmas \cite{gokhale_evolutionary_2010}.

In this Letter, we introduce a general framework to extend social dilemmas to  structured populations with the presence of interactions in groups of variable size. 
We model the interactions structure of a population of players as a hypergraph where players are involved in both pairwise and higher-order games represented as hyperedges of different sizes. In this way, the payoff of each player is determined by the strategies of all the players involved in the interaction at once.
By comparing extensive numerical simulations of the evolutionary dynamics of the game on random hypergraphs to the analytical results in well-mixed approximation, we find that the higher-order interactions can dramatically change the Nash Equilibria (NE) of the game, allowing cooperators to survive in the Prisoner's Dilemma (PD).
In fact, above a critical number of higher-order interactions which depends on the parameters of the game, the dynamics shows an explosive transition to a bistable state, where besides full defection (the only NE in case of just pairwise interactions) a cooperative stable state emerges. 
Moreover, we found that an initial critical mass of cooperators is also needed to sustain cooperation in the long term: below this critical mass every player becomes a defector, even if the number of higher-order interactions is above the critical threshold.

\paragraph*{The model.} We consider a population of $N$ players taking part in $M$ strategic interactions, which can either be pairwise or in groups of three or more players. Such interactions are described by a hypergraph $\mathcal{H}(\mathcal{V}, \mathcal{E})$, where $\mathcal{V}$ is the set of $N$ vertices or nodes representing players, and $\mathcal{E}$ is the set of $M$ hyperedges \cite{battiston_networks_2020, battiston_physics_2021}.
Each hyperedge $e_g$, with $g \in 1, \cdots, M$, is a group (a subset of $\mathcal{V}$)  of two or more players interacting in game $g$.  
The hypergraph can be represented by a $N \times M$ incidence matrix $B$, whose entry $b_{i g}$ is equal to $1$ if the player $i$ is playing the game $g$, or is zero otherwise. 
The number of games in which a player $i$ takes part is given by the hyperdegree $ k_i = \sum_{g = 1}^M b_{i g} $, while the number of players in a game $g$ is the size of the hyperedge $q_g = |e_g| = \sum_{i = 1}^N b_{i g}$. 
We focus here on the case of hypergraphs with hyperedges of size two (2-hyperedges, or simply edges) and three (3-hyperedges), respectively corresponding to classical pairwise games (2-games) and games played in groups of three players (3-games).
\begin{figure}[htp!]
    \centering
    \includegraphics[width=0.47\textwidth]{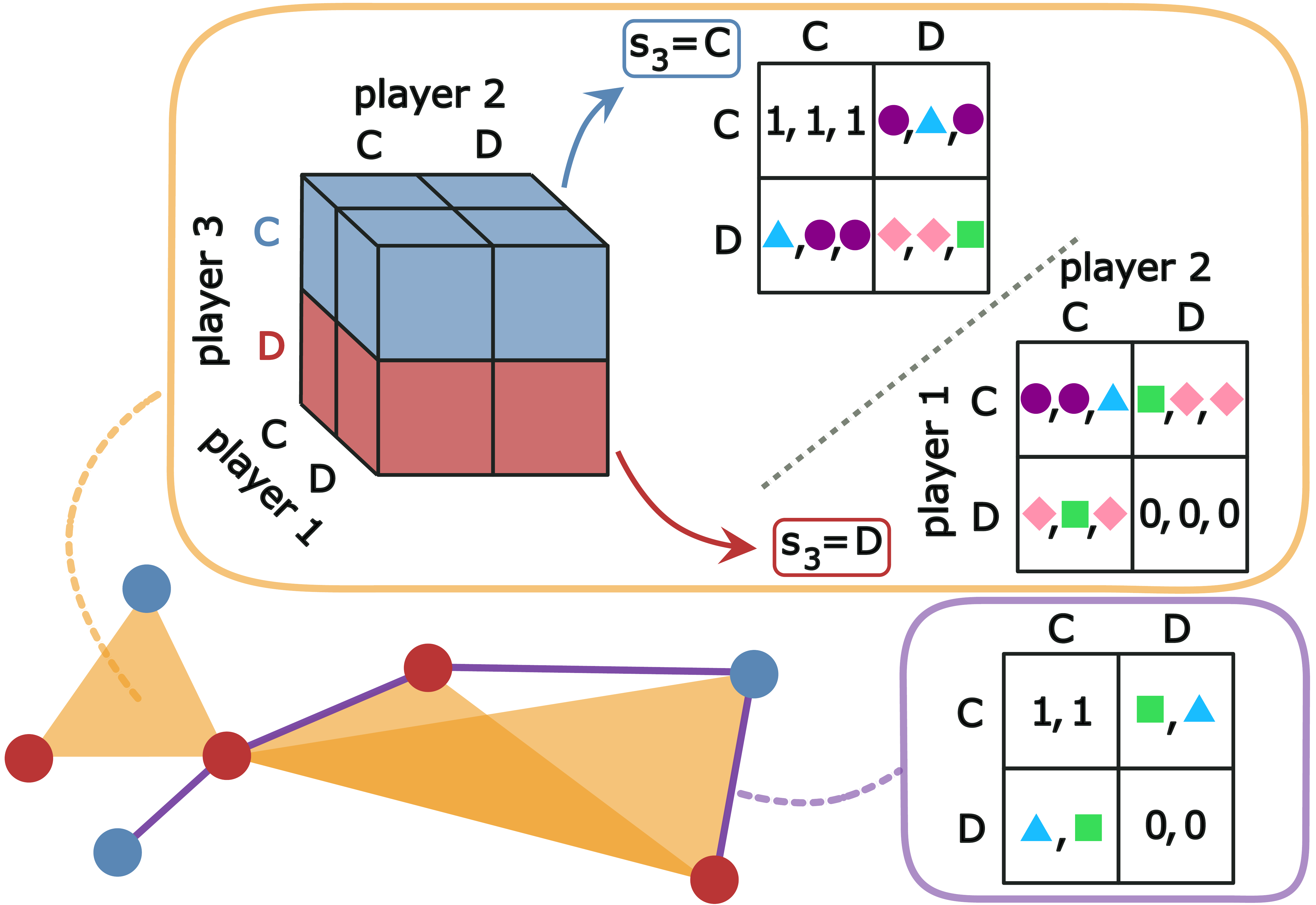}
    \caption{Higher-order games on a hypergraph. 
    The orange triangular areas are hyperedges of size $q_g=3$, corresponding to
    games played by three players 
    (3-games), while the purple segments are pairwise interactions, hyperedges of size $q_g=2$, representing standard games played by two players (2-games). The payoff structures of symmetric 2-games and 3-games are reported in the two boxes.}
    \label{fig:schematic}
\end{figure} 
Concerning the payoffs, given that we have $q_g$ players involved in a game $g$, if we indicate as $n_s$ the number of different strategies available, the identical players (or symmetry) requirement reduces the total number of different payoffs $n_s^{q_g}q_g$ to just $n_s\binom{n_s + (q_g - 1) -1}{(q_g - 1)}$ payoffs (see SM). As in the classical pairwise symmetric 
games, here we consider only $n_s=2$ possible strategies, cooperation (C) and defection (D). 
This means that for symmetric 2-games there are 4 possible different payoffs while for 3 players there are 6.
As usual, the payoffs for 2-games can be displayed as a $2 \times 2$ matrix $\Pi$, whose element $\pi_{s_i s_j} = \left[\pi_{s_i}(s_j),\pi_{s_j}(s_i)\right]$ is the pair of payoffs for player $i$ and $j$ respectively, when the first player plays strategy $s_i$ and the second $s_j$.
Generalizing this approach to the case of interactions in groups of three players,  
the payoffs for a 3-game can then be represented as a $2 \times 2 \times 2$ tensor $\mathcal{T}$, whose element $\tau_{s_i s_j s_k} = \left[\tau_{s_i}(s_j, s_k),\tau_{s_j}(s_i, s_k), \tau_{s_k}(s_i, s_j)\right]$ is now a 3-tuple with the value of the payoff for each of the three players $i$, $j$ and $k$, playing strategies $s_i$, $s_j$, $s_k$. 
Fig.~\ref{fig:schematic} illustrates how to implement social dilemmas on higher-order systems. 
The complete payoff structure for both 2-games ($q_g=2$) and 3-games ($q_g = 3$) is shown, using different symbols to denote different values of payoffs. 
As commonly done in the study of social dilemmas, without loss of generality we choose  the payoff for mutual cooperation equal $1$, while the payoff for mutual defection is equal $0$, for both 2-games and 3-games \cite{guo_evolutionary_2021}. 
In a similar manner, i.e. independently from the number of players (2 or 3) in the game, with \Tpayoff and \Spayoff we indicate 
the  payoffs received for unilaterally deviating from mutual cooperation and defection respectively. In this way it is immediate to identify in \Tpayoff and in \Spayoff the payoffs usually denoted, in pairwise social dilemmas, as the \emph{temptation} $T$ and the \emph{sucker}'s payoff $S$.
Identifying $T$ and $S$ is a fundamental step for the characterization of the game. According to the values of $T$ and $S$, classical pairwise games are classified into four different types, each characterized by a different set of Nash Equilibria (NE): 
the Prisoner's Dilemma ($T>1$, $S<0$), the Chicken game ($T>1$, $S>0$), the Stag Hunt game ($S<0$, $T<1$) and the Harmony game ($S>0$, $T<1$) (see SM). Hence, we can now extend the same classification to 3-games.
In this case there are two additional payoffs, namely for defection against a cooperator and a  defector ($\Wpayoff$ namely W), and for  cooperation against a cooperator and a defector ($\Gpayoff$ namely G). According to the relative value of these two additional payoffs (if $G>W$ or $G<W$) each type of 3-games is divided in two disjoint subsets with different Nash Equilibria (see SM).  

\paragraph*{Results.} 

\begin{figure}[htp!]
    \centering
    \includegraphics[width=0.47\textwidth]{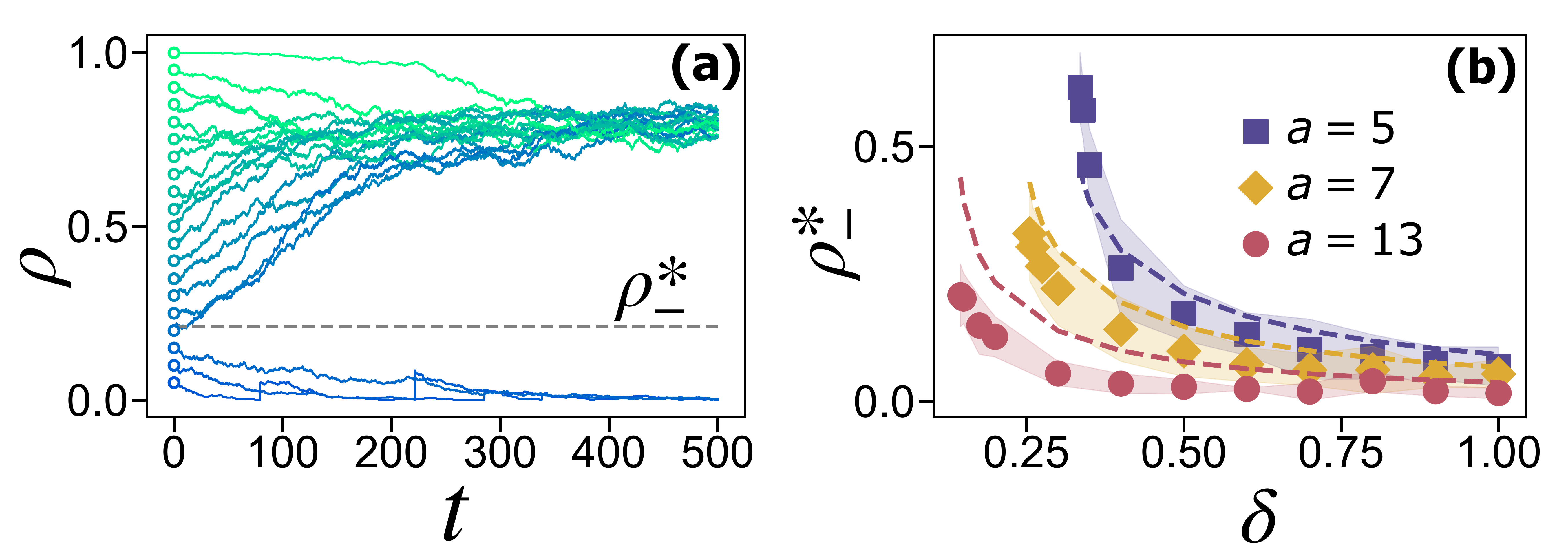}
    \caption{Basins of attraction and critical mass of cooperators for the PD on random hypergraphs. 
    \textbf{(a)} Temporal evolution of the fraction of cooperators for various initial conditions and $\delta=0.5$, $a=5$, $\langle k \rangle=20$.
    \textbf{(b)}  Unstable stationary state $\rho^*_-$ as a function of $\delta$ for average hyperdegree $\langle k \rangle=20$ and different values of $a$. Symbols show the numerical results, while the dashed lines are the analytical mean-field predictions. The shaded areas represent the errors.} 
    \label{fig:unstable}
\end{figure}

To investigate the effects of higher-order interactions on the equilibria of a system with $N$ players, we considered the following evolutionary game dynamics. We start from a population with an initial fraction $\rho_0 = \rho(t=0)$ of cooperators. At each time step, one player (namely the focal) is selected at random, and a second player (the model) is chosen at random among the neighbouring nodes of the focal player on the hypergraph, i.e. those nodes which are connected to the focal player by hyperedges of any size. 
Each of the two selected players plays a 2-game with all its neighbors connected through a 2-hyperedge, and a 3-game for each 3-hyperedge it takes part in.
A 2-game is completely defined by the values of the payoff matrix entries $T$ and $S$, while the 3-game has the same $T$ and $S$ of the 2-game, but is also defined by the payoffs $G$ and $W$.
For each game, the focal (respectively model) player earns a payoff depending on its strategy and on the strategies of the other players involved in that particular game (i.e. of the other players in the hyperedge). The sum of all the game payoffs defines the total payoff $\pi_f$ of the focal player and the total payoff $\pi_m$ of the model player. 
The focal player has then the possibility to adopt the strategy of the model player $s_m$, with a probability which is a non-decreasing function of the total payoff difference $\pi_m - \pi_f$, modelled as a Fermi function~\cite{blume_stat_strat_1993, szabo_evolutionary_1998, traulsen_EGTcooperation_2006}:
   $ p_{s_f \to s_m} = \{1 + exp[-w(\pi_m - \pi_f)]\}^{-1}$
where $w$ represents the strength of selection. 
Since we are interested in the Nash Equilibria of the game we iterate the dynamics to compute 
the quasistationary (QS) probability distribution \cite{de_oliveira_how_2005, sander_sampling_2016}  of the fraction of players adopting strategy $C$ (cooperators), whose maxima correspond to the Evolutionary Stable States  (ESS) \cite{taylor_evolutionary_1978}, $\rho^*$, of the evolutionary dynamics \cite{zhou_evolutionary_2010, faure_quasi-stationary_2014, civilini_evolutionary_2021}.
As for the underlying structure of interactions, we have considered random hypergraphs with different numbers  of higher-order interactions. We have constructed hypergraphs of order $N$ with tunable average hyperdegree $\langle k \rangle = \sum_{i=1}^N k_i / N$ and fraction of 3-hyperedges $\delta = {n_{\Delta}}/{ M }$, 
where $M=n_{\Delta} + n_{/}$ is the sum of the total number of $n_{/}$ 2-player and $n_{\Delta}$ 3-player interactions in the hypergraph (see SM).
\begin{figure}[htp!]
    \centering 
\includegraphics[width=0.47\textwidth]{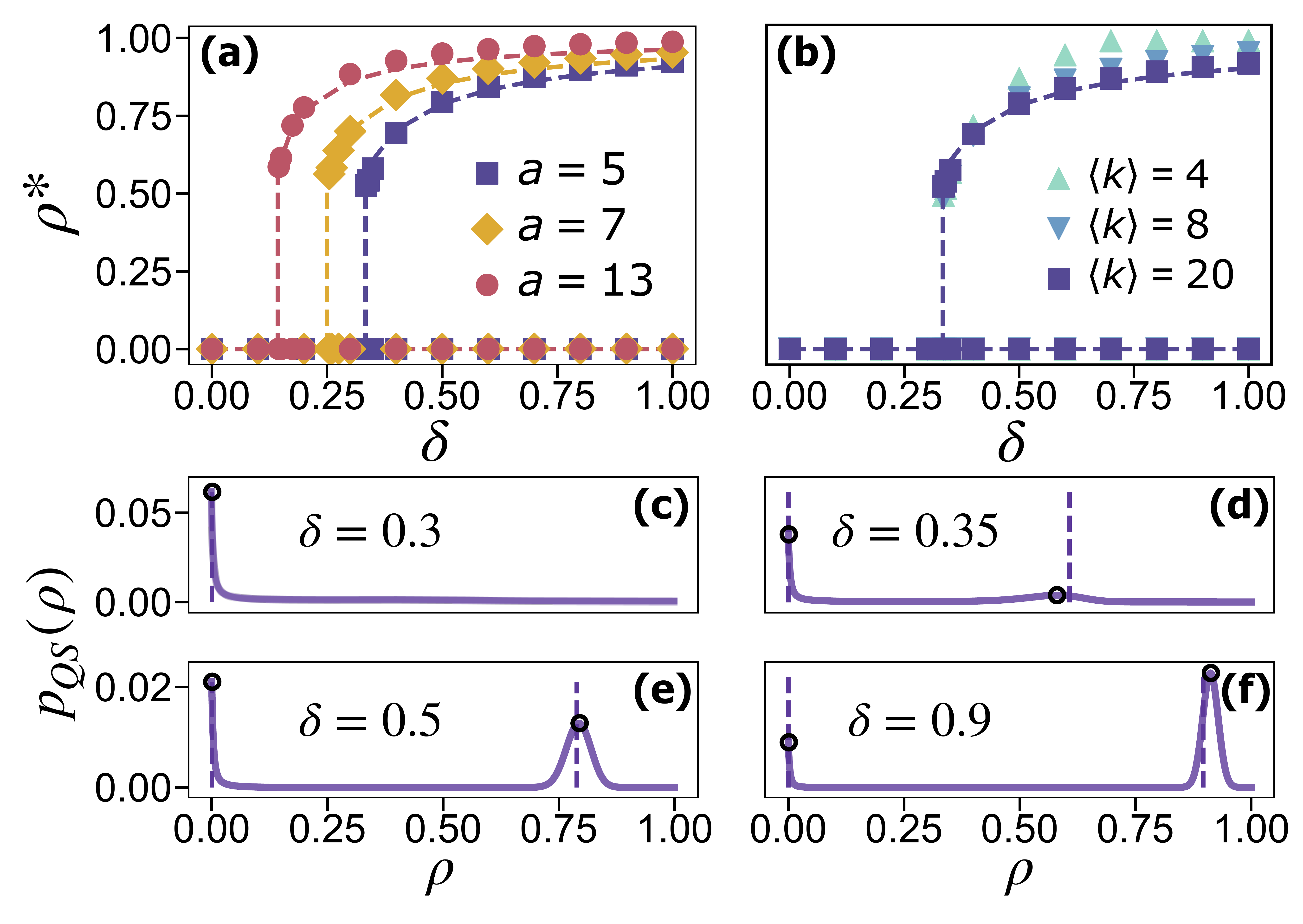}
    \caption{Stable stationary states for the PD on random hypergraphs with $N=1000$ and tunable ratio $\delta$ of three-body interactions. \textbf{(a)} Fraction of cooperators at equilibrium as a function of $\delta$ for average hyperdegree $\langle k \rangle=20$ and different values of $a$, and \textbf{(b)} for $a=5$ and different values of $\langle k \rangle$.
\textbf{(c-f)} Quasistationary distributions for $a=5$ and $\langle k \rangle=20$ and four different values of $\delta$. Symbols represent the numerical results averaged over 1000 independent runs (the error bars are smaller than the symbols), while dashed lines are the analytical mean-field predictions.}
\label{fig:stable}
\end{figure} 
Fig.~\ref{fig:stable} shows the results for the case of the Prisoner Dilemma  (PD), the most  relevant game in the study of social dilemmas. We recall that the pairwise PD is defined by payoff values $T>1$ and $S<0$. In particular, for our simulations we chose $T=1.5$, $S=-0.5$ and for the strength of selection $w=1/\langle k \rangle$ (see SM). For the 3-game PD we consider $G$ and $W$ such that $(G - W) > 0$, since in this case the one-shot 3-game has 4 different NE: full defection (D,D,D) and all the permutations of 2 cooperators and 1 defector (see SM). 
In the two top panels we show $\rho^*$, the ESS of the Replicator dynamics (RD), as  a function of the fraction $\delta$ of 3-hyperedges, for different values of $a := 2(G-W)$ and of $\langle k \rangle$, the average hyperdegree of the hypergraph. The colored symbols represent the numerical results for $\rho^*$, obtained from the peaks of the QS distribution $p_{QS}(\rho)$ in panels (c-f). 
We observe a bifurcation in the stable points of the dynamics when the fraction $\delta$ of 3-hyperedges exceeds a critical value $\delta_c(a)$.  
In particular, while for $\delta<\delta_c$ the only stable NE is full defection $\rho^*_D = 0$, as in the standard pairwise PD,
for $\delta>\delta_c$ 
we observe the emergence of a bistable behaviour where cooperation survives: 
besides the full defection $\rho^*_D$, a new stable state $0.5 \leq \rho^*_+ \leq 1$ appears due to the effect of the higher-order interactions. 
Fig.~\ref{fig:unstable}(a) illustrates the typical time evolution of the system. It reports the fraction of cooperators $\rho(t)$ as a function of time for 20 different initial conditions characterized by different initial values $\rho_0 = \rho(0)$.  
We notice that when $\rho_0$ is smaller than a given threshold $\rho^*_-$, the dynamics will converge to the full defection state. Conversely, when 
$\rho_0 > \rho^*_-$, it will converge to the stable state $ \rho^*_+$ where a finite fraction of the population are  
cooperators. In other words, $\rho^*_-$ represents the initial critical mass of cooperators needed for cooperation to survive in the long term. 
Fig.~\ref{fig:unstable}(b) shows  that $\rho^*_-$ is a decreasing function of $\delta$ 
for any value of the parameter $a$. 
This implies that smaller initial densities of cooperators are sufficient to sustain stable cooperation in systems with a larger fraction $\delta$ of 3-games interactions (see SM for further details on the numerical results).

To better characterize the role of higher-order interactions on the outcome of the game,  
we have solved analytically the game in the well-mixed population case, where each player can interact with all the others. Each interaction is considered to be a 3-game, with probability $\delta$, and a 2-game with a probability $1-\delta$.
The evolutionary dynamics of the fraction $\rho$ of cooperators for a well-mixed population in the thermodynamic limit is given by the mean-field Replicator Equation (RE) \cite{hofbauer_evolutionary_1998, fokkerplanckmastereq1, traulsen_coevolutionary_2006}: 
\begin{equation} \label{eq:replicator_eq}
    \frac{d\rho}{dt} = \rho(1-\rho)\left[\pi_C(\rho,\delta) - \pi_D(\rho,\delta)\right]
\end{equation}
where $\pi_C$ and $\pi_D$ are respectively the expected payoff of a cooperator and of a defector and are both functions of the density of cooperators $\rho$ and of the fraction $\delta$  of 3-games (see SM).
Hence, the expected payoff difference is also a function of $\rho$ and $\delta$:
\begin{equation}
    \pi_C - \pi_D = -\rho^2c \delta  + \rho\left(c \delta - b- 2S\right) + S
\end{equation}
where $a=2(G-W)$, $b=T-S-1$ and $c=(a+b)$. 
Therefore, besides the two trivial stationary absorbing states of the RE, namely full-defection $\rho^*_D = 0$ and full cooperation $\rho^*_C = 1$, Eq.~\ref{eq:replicator_eq} has other two stationary states $\rho^*_{\pm}$ for which $\pi_C - \pi_D = 0$:
\begin{equation}
    \rho^*_{\pm} = \frac{c \delta - b - 2S \pm \sqrt{ (c \delta - b)^2 + 4S(b+S)}}{2c\delta}
\end{equation}
It follows immediately that when $\Delta = \left( c \delta - b \right)^2 + 4S(b+S) \geq 0$, then $\rho^*_{\pm}$ are real-valued for every $a, b, \delta, S$. In particular, given that $\left(c \delta - b \right)^2$ is always positive, a sufficient condition for the existence of real-valued solutions is $4S(b+S)=4S(T-1)>0$, which is always satisfied for the Stag Hunt game and Chicken game.
For the Prisoner's Dilemma and the Harmony game instead $\Delta > 0$ holds only for certain values of the parameters.
In particular, for the game we are focusing on in this Letter, namely the PD, we have $T>1$ and $S<0$, hence $b = T - S - 1>0$. 
Moreover, $c=a+b>0$, given that we are considering 3-games with $a=2(G-W)>0$.
In this case, we find that $\rho^*_{\pm}$ are real-valued for:
\begin{align}
    \delta >& \delta^{\textrm{th}}_1= \frac{b+\sqrt{-4S(b+S)}}{c} \label{eq:sol_existence}
    \\
    \delta <& \delta^{\textrm{th}}_2= \frac{b-\sqrt{-4S(b+S)}}{c}
\end{align}
It is easy to prove (see SM) that if $\delta < \delta^{\textrm{th}}_2$ the real-valued solutions $\rho^*_{\pm}$ are negative, while if  $\delta > \delta^{\textrm{th}}_1$, $\rho^*_{\pm}$ are always positive and such that  $ 0 \leq \rho^*_- \leq 0.5 \leq \rho^*_+ \leq 1$.
The stability analysis of the solutions yields that,  while $\rho^*_D=0$ and $\rho^*_{+}$ are stable, $\rho^*_{-}$ and $\rho^*_C=1$ are unstable stationary states.
Therefore, Eq.~\eqref{eq:sol_existence} gives us the mean field critical threshold $\delta^{\textrm{th}}_1$ of 3-player interactions for cooperation to survive in the higher-order PD. In fact, if the number of 3-player interactions is below this critical threshold the only stable stationary state of the higher-order PD is full defection $\rho^*_D=0$, as in the pairwise case. If instead the fraction of 3-games $\delta$ exceeds $\delta^{\textrm{th}}_1$, an explosive transition to a bistable state emerges, where both $\rho^*_D = 0$ and $0.5 \leq \rho^*_+ \leq 1$ are stationary stable states. 
In Fig.~\ref{fig:stable} the analytical mean-field results are reported as dashed lines. In particular, the analytical predictions for the stable states $\rho^*_+$ and $\rho^*_D$  are in perfect agreement with the peaks of the quasistationary distributions in Fig.~\ref{fig:stable}(c-f) and with the symbols in panels (a-b) reporting the ESS obtained numerically on random hypergraphs. At the same time, also the critical fraction of 3-games $\delta^{\textrm{th}}_1$ (vertical dashed lines in Fig.~\ref{fig:stable}(a-b) ) correctly marks the discontinuous transition to bistability observed numerically.
Fig.~\ref{fig:unstable} displays the unstable solution $\rho^*_-$, which defines the basins of attraction of the two stationary stable states $\rho^*_D$ and $\rho^*_+$, showing again a good agreement between the  mean-field predictions (dashed lines) and the numerical results (trajectories in Fig.~\ref{fig:unstable}(a) and symbols in Fig.~\ref{fig:unstable}(b)).
\paragraph*{Conclusions.}
In this Letter, we have introduced 
a general game theory framework to study social dilemmas in systems where not only pairwise but also higher-order interactions are possible.  
The main finding of our work is 
that cooperation can survive even in cases, such as the PD, where pairwise interactions would lead to full defection. Moreover, 
the observed transition to a state 
with a stable fraction of cooperators is explosive when the number of higher-order interactions of the system is above 
a critical threshold that depends on the parameters of the game. 
However, the observed bistability implies that, even when possible, the survival of cooperators is not guaranteed: a critical mass of initial cooperators is in this case  needed to achieve stable pro-social behaviour.
This is in agreement with empirical observations regarding the critical mass of initiators required to trigger social and cultural changes \cite{centola_experimental_2018, pereda_group_2019}.
Our findings demonstrate that higher-order interactions  can promote cooperation in competitive environments,  showing a new way out of  social dilemmas. 
While in this Letter we have been focusing on the PD, our higher-order framework can be easily applied to any other game. So we hope, our work will inspire new research on the investigation of higher-order interactions and their effects in different strategic scenarios.

\bibliographystyle{apsrev4-2} 
\bibliography{main}

\cleardoublepage
\newpage
\onecolumngrid

\setcounter{figure}{0}
\setcounter{table}{0}
\setcounter{equation}{0}
\makeatletter
\renewcommand{\thefigure}{S\arabic{figure}}
\renewcommand{\theequation}{S\arabic{equation}}
\renewcommand{\thetable}{S\arabic{table}}

\setcounter{secnumdepth}{2} 

\section*{\large{Supplemental Material: Explosive cooperation in  social dilemmas on higher-order networks}}
\normalsize
\vspace*{0.2 cm}

\setcounter{secnumdepth}{4}

\section*{Number of different payoffs}

We consider a general $q_g$-person game, where each player can choose among $n_s$ different strategies.
If the players are distinguishable (i.e. not identical) then there are $n_s^{q_g}$ possible different elements of the payoff tensor and for each element, there are $q_g$ possible different individual payoffs (since each player is different). That is, in the case of distinguishable players the maximum number of different payoffs $N^{max}_{\pi}$ is:
\begin{equation}
    N^{max}_{\pi} = n_s^{q_g}q_g
\end{equation}
Instead, if the players are identical, the payoff of a player depends on its strategy and on the unordered sample of the strategies of the other $q_g - 1$ players. Unordered because, since the players are identical, it does not matter which player plays which strategy. Given that there are $n_s$ possible different strategies, by applying the formula for unordered sampling with replacement of $q_g-1$ items picked at random from $n_s$ choices, we find that for identical players (i.e. for symmetric games) the maximum number of different payoffs is:
\begin{equation}
    N^{max}_{\pi} = n_s\binom{n_s + (q_g - 1) -1}{(q_g - 1)}
\end{equation}
Substituting, $q_s=3$ and $n_s=2$, we get $N^{max}_{\pi} = 6$. In our model we set the payoff for mutual cooperation $R=1$ and that for mutual defection $P$ equals $0$. The remaining four payoffs are then denoted as \Gpayoff, \Wpayoff, \Tpayoff, and \Spayoff.

\section*{Classification of social dilemmas}

\begin{figure}[htp!]
  \centering
  \includegraphics[width=0.3\linewidth]{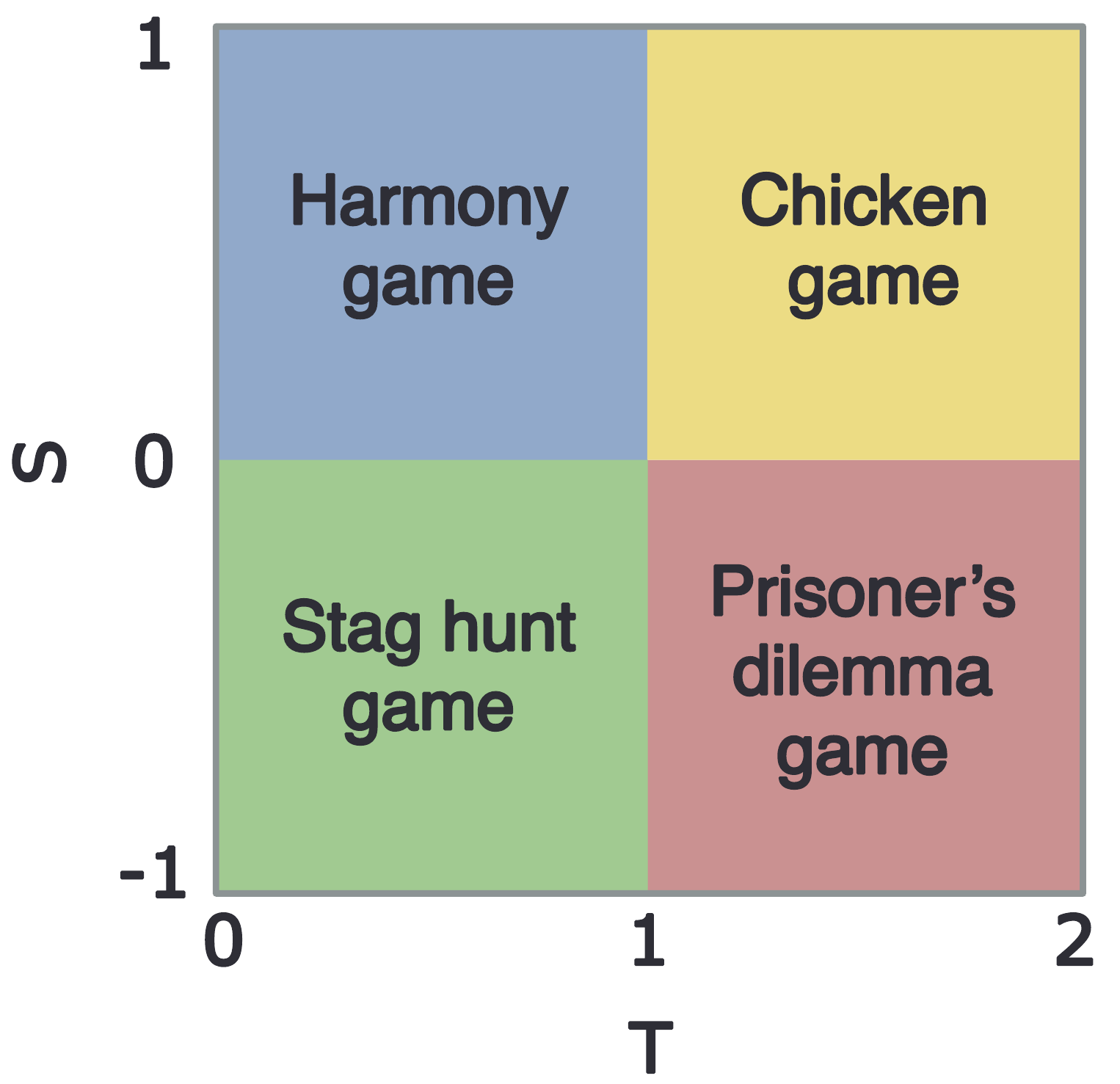}
  \caption{Games square showing the four different games defined by the combination of values of $T$ and $S$ above or below the thresholds $R=1$ and $P=0$.}
\label{fig:games_matrix}
\end{figure}

In a social dilemma each player can choose between two strategies, either to cooperate (strategy $C$) or to defect (strategy $D$) \cite{szabo_evolutionary_1998, szabo_evolutionary_2007,  doebeli_evolutionary_2004}.
Defecting brings a higher individual payoff than cooperating when facing one or more cooperators (i.e. it is convenient to free-ride the cooperative efforts of the other players). However, if all players defect, everyone (defectors included) suffers, since the collective payoff vanishes.
For pairwise (i.e. 2-player) social dilemmas, the payoffs for mutual cooperation (namely Reward, $R$) and mutual defection (namely Penalty, $P$) can be set to 1 and 0 respectively, without loss of generality.
Moreover, the payoff associated with unilaterally deviating from mutual cooperation is $T$ (Temptation), while a player receives the payoff $S$ (Sucker) for deviating from mutual defection.
It follows that if $S > 0$ it is convenient for a rational player to deviate from mutual defection, while if $T > 1$ it is preferable to deviate from mutual cooperation.
Therefore, depending on the combination of values of $T$ and $S$ (i.e, above or below the threshold values $1$ and $0$), we get four scenarios that depict four possible different games. These games are characterized by different Nash equilibria and can be conveniently represented as a \textit{square of games} as shown in Fig.~\ref{fig:games_matrix}. In particular, we have a Prisoner's Dilemma for $T>1, S<0$, a Chicken game for $T>1, S>0$, a Stag Hunt game for $T<1, S<0$, and $T<1, S>0$ define a Harmony game.

\begin{figure}[htp!]
  \centering
  \includegraphics[width=0.5\linewidth, valign=t]{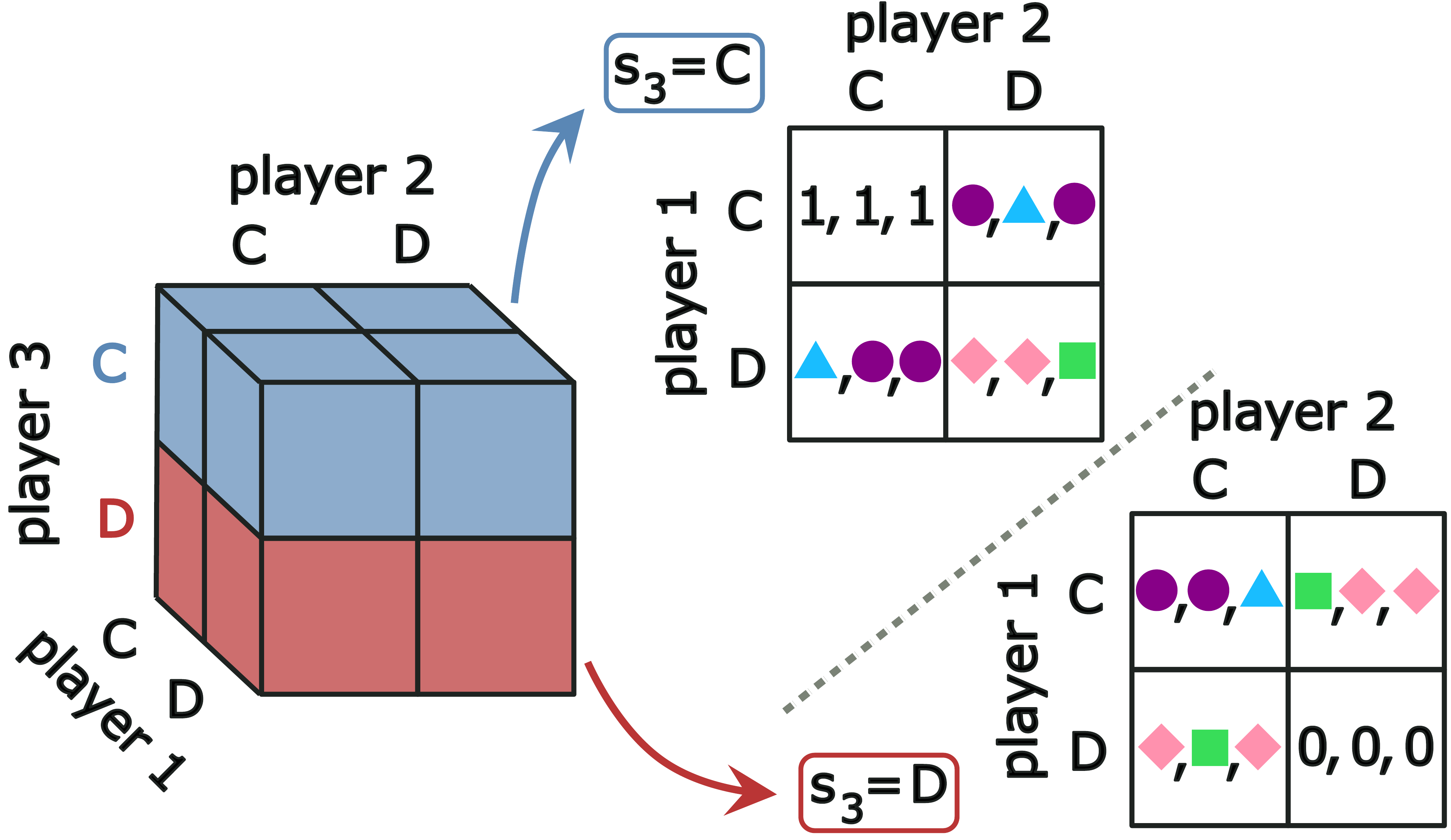}
  \caption{Payoff tensor for 3-player social dilemmas. Consistently with the notation for pairwise games, we assume the payoffs for mutual cooperation and defection respectively equal to $(1,1,1)$ and $(0,0,0)$. The two matrices represent the two levels of the $2\times 2 \times 2$ payoff tensor, whose elements are the triplets of payoffs $(\tau_1, \tau_2, \tau_3)$. The top matrix shows the payoffs when player 3 adopts cooperation, i.e. for $s_3 = C$. The matrix on the bottom reports the payoffs for the case $s_3 = D$. It is worth noticing that despite the game being symmetric it would be not obvious to reconstruct the whole payoff tensor just from the payoffs of player 1, as usually done in the case of pairwise symmetric games.}
\label{fig:payoff_tensor_colors}
\end{figure}

A pairwise game can be represented using the so-called payoff matrix representation, where the element of the matrix $\pi_{s_i s_j} = \left[\pi_{s_i}(s_j),\pi_{s_j}(s_i)\right]$ is the pair of payoffs for player $i$ and $j$ respectively, when the first player plays strategy $s_i$ and the second $s_j$ \cite{weibull_evolutionary_2004}.
For 3-player games, the payoff matrix is then substituted by a $2\times 2 \times 2$ payoff tensor $\mathcal{T}$, whose element $\tau_{s_i s_j s_k} = \left[\tau_{s_i}(s_j, s_k),\tau_{s_j}(s_i, s_k), \tau_{s_k}(s_i, s_j)\right]$ is now a 3-tuple with the value of the payoff for each of the three players $i$, $j$ and $k$, playing strategies $s_i$, $s_j$, $s_k$. 
As seen in the first section of the SM, the number of possible different payoffs for 3-player symmetric games is equal to 6.  Consistently with the pairwise social dilemmas, we choose the payoff for full cooperation (i.e. strategy profile $(C,C,C)$) equal to $(1,1,1)$ and the payoff for mutual defection (strategy profile $(D,D,D)$) equal to $(0,0,0)$.
As shown in the manuscript, in a 3-player game the payoff $\Tpayoff$  for unilaterally deviating from mutual cooperation (respectively the payoff $\Spayoff$ for deviating from mutual defection) is analogous to the temptation payoff $T$ (respectively sucker's payoff $S$) in the pairwise social dilemma. 
Figure \ref{fig:payoff_tensor_colors} shows the full $2\times 2 \times 2$ payoff tensor.
In fact, as in pairwise games, in 3-person games, it is advantageous for a rational player to deviate from mutual cooperation if $\Tpayoff>R=1$, while it is beneficial to deviate from mutual defection if $\Spayoff>P=0$. 
However, unlike the pairwise games, in 3-person games there are two additional payoffs ($\Wpayoff$ and $\Gpayoff$) that define a new threshold. 
In particular, if $\Wpayoff > \Gpayoff$, it is favorable to be a defector when playing against a cooperator and a defector, while if $\Wpayoff < \Gpayoff$, it is convenient to side with the cooperator. 

\subsection*{Classification of 3-player games}
We saw that, given their definitions, the payoffs \Tpayoff and \Spayoff can be effectively regarded as the 3-player extension of $T$ and $S$, and hereafter to avoid confusion we will refer to them with the same letters of the pairwise case (i.e. \Tpayoff as $T$ and \Spayoff as $S$).
Hence, we can extend to 3-games the same classification based on the values of $T$ and $S$ of pairwise social dilemmas. That is, for the 3-player games, according to the values of $T$ and $S$ we have the four social dilemmas we saw for the pairwise case. However for 3-person games, depending on whether $\Gpayoff>\Wpayoff$ or $\Gpayoff<\Wpayoff$, each of these four games is now further divided into two disjoint subsets with different NE as shown in the following list.

\subsubsection*{3-player Prisoner's Dilemma game}
Pairwise Prisoner's Dilemma (PD) is defined by $T > 1$ and $S < 0$.
For 3-player, the condition $\Gpayoff > \Wpayoff$ defines two PD with different NE:
\begin{itemize}
    \item $\Wpayoff > \Gpayoff$:  $(D,D,D)$ is the only NE of the game.
    \item $\Wpayoff < \Gpayoff$: the game has $4$ different pure NE, $(D, D, D)$, $(C, C, D)$, $(C, D, C)$, $(D, C, C)$.
\end{itemize}

\subsubsection*{3-player Harmony game}
Pairwise Harmony games are defined by $T < 1$ and $S > 0$.
Depending on the values of $\Gpayoff$ and $\Wpayoff$ we now have the following Nash Equilibria:
\begin{itemize}
    \item $\Wpayoff > \Gpayoff$:  the game has $4$ different pure NE, $(C, C, C)$, $(C, D, D)$, $(D, C, D)$, $(D, D, C)$.
    \item $\Wpayoff < \Gpayoff$:  $(C,C,C)$ is the only NE of the game.
\end{itemize}

\subsubsection*{3-player Chicken game}
The pairwise Chicken game (CG) is defined by $T > 1$ and $ S > 0$.
The values of $\Wpayoff$ and $\Gpayoff$ characterize two different subsets of CG with different Nash Equilibria as:
\begin{itemize}
    \item $\Wpayoff > \Gpayoff$: the NE are $(D,D,C)$, $(D,C,D)$ and $(C,D,D)$.
    \item $\Wpayoff < \Gpayoff$: the NE are $(C,C,D)$, $(C,D,C)$ and $(D,C,C)$. 
\end{itemize}

\subsubsection*{3-player Stag hunt game}
The Stag hunt game is defined by $T < 1$ and $S < 0$.
In this case, the values of the payoffs $\Wpayoff$ and $\Gpayoff$ do not change the two NE, $(C,C,C)$ and $(D,D,D)$.
However, depending on which payoff between $\Gpayoff$ and $\Wpayoff$ is higher, the ways in which is possible to reach these two NE changes, and  one NE is favored over the other. In Game theory notation, this threshold influences the basin of attraction of the two NE, without changing the NE themselves, i.e. it makes one or the other NE risk dominant:
\begin{itemize}
    \item $\Wpayoff > \Gpayoff$: there are more strategic moves leading to $(D,D,D)$ (it has a larger basin of attraction, i.e. it is risk dominant) than to $(C,C,C)$; defection is promoted over cooperation. 
    \item $\Wpayoff < \Gpayoff$: cooperation is promoted since there are more strategic paths bringing to $(C,C,C)$.
\end{itemize}

\section*{Generating random hypergraphs}

\begin{figure}[htp!]
    \centering
    \includegraphics[width=\columnwidth]{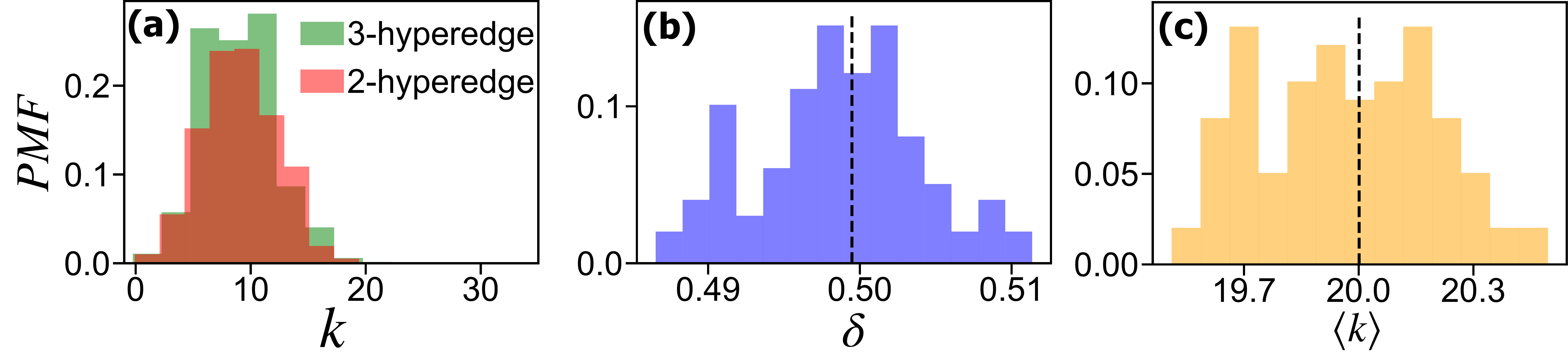}
    \caption{\textbf{(a)} Probability mass function (PMF) of the hyperdegree $k$, distinguishing between the contribution to $k$ of 2-hyperedges and 3-hyperedges. \textbf{(b)} PMF of the fraction of 3-hyperedges $\delta$. \textbf{(c)} PMF of the average hyperdegree $\langle k \rangle$ of an hypergraph. The PMF are computed over 100 instances of a random hypergraph of size $N=1000$. The dotted lines denote the mean of the distributions. The desired fraction of 3-hyperedges and average hyperdegree are $\delta = 0.5$ and $\langle k \rangle = 20$, respectively. }
    \label{fig:hyperdeg_dist}
\end{figure}

As for the underlying structure of interactions, we considered random hypergraphs with  different numbers  of higher-order interactions. We have constructed hypergraphs with tunable numbers $n_{/}$ and $n_{\Delta}$ of 2- and 3-hyperedges, respectively.
Let  $\delta = {n_{\Delta}}/{ M }$, where $M=n_{\Delta} + n_{/}$ is the total number of interactions in the hypergraph, be the fraction of 3-hyperedges. For fixed values of $N$, $\delta$ and $M$, we start with $N$ nodes and first connect each of the possible $N(N-1)/2$ pairs of distinct nodes with a probability 
\begin{equation}
    p_/ = (1- \delta ) \langle k \rangle/ (N-1)
\end{equation}
where $\langle k \rangle = M / N $ is the desired average hyperdegree of each node. 
We then connect each of the $N(N-1)(N-2)/6$ triplets of distinct nodes with a probability 
\begin{equation}
  p_{\Delta} = 2   \delta \langle k \rangle /((N-1)(N-2))
\end{equation}
Given that every time we add a pairwise interaction the total hyperdegree of the network increases by $2$, while when we add a 3-hyperedge it increases by $3$, we obtain a random hypergraph with the desired $\langle k \rangle$ and $\delta$. If the final hypergraph is not connected, we take the largest connected component. Fig.~\ref{fig:hyperdeg_dist} shows that the networks obtained from this algorithm correctly reproduce the desired numbers of 2-hyperedges, 3-hyperedges, and average hyperdegree. In particular, in Fig.~\ref{fig:hyperdeg_dist}.a we notice that the numbers of 2-hyperedges and 3-hyperedges in which each player takes part in are distributed as binomial distributions centered around $k = 10$, as expected for random hypergraphs given the chosen parameters $\langle k \rangle = 20$ and $\delta = 0.5$ \cite{newman_networks_2010}. 

\section*{Details of the numerical results}

\subsection*{ Stable states }
In order to estimate $\rho^*_+$, we simulated $1000$ runs of the evolutionary dynamics. For each run, we start with a randomly chosen fraction of defectors $0<\rho_0 = \rho(t=0)<1$ and we use a different instance of the random hypergraph (generated using the algorithm described in the SM). We use the quasistationary (QS) method \cite{de_oliveira_how_2005, sander_sampling_2016} to evolve the system allowing sufficient time for thermalization. In particular, for our simulations on hypergraphs of size $N=1000$, we chose a thermalization time of $10^6$ time steps and a total simulation length of $10^7$ time steps. We recall from the manuscript that we are focusing on the case of the Prisoner's Dilemma and to define the game we arbitrarily chose the payoff values $T=1.5$ and $S=-0.5$.
As for the strength of selection we chose $w=1/\langle k \rangle$, however we have verified that the results are consistent for at least one order of magnitude above and below this choice for $w$. We chose $w$ proportional to $1/\langle k \rangle$ in order to have a comparable strength of selection among different hypergraphs with different $\langle k \rangle$, since the average payoff in the hypergraph increases with the average hyperdegree.
It has been proved that the peaks of the QS probability distribution of players with a given strategy (in our case cooperators) correspond to the evolutionary stable state (ESS)  of the system \cite{zhou_evolutionary_2010}. We, therefore, find the peak(s) of the QS probability distribution obtained by averaging the distribution of cooperators over all the 1000 runs. 
Through this method, if $\delta < \delta_c$ we obtain one peak (corresponding to $\rho^*_D$) of the QS distribution, instead if $\delta > \delta_c$ we find two peaks ($\rho^*_D$ and $\rho^*_+$) . 
To estimate the error on our measurements of the ESS, we first found the peaks of the QS probability distribution of each of the 1000 runs, obtaining in this way one peak, $(\rho^*_D)_i$, or two peaks, $(\rho^*_D)_i$ and $(\rho^*_+)_i$, for each run $i$. We then compute the absolute deviations of these peaks from the measured ESS (i.e., from the corresponding peak of the QS distribution averaged over all 1000 runs). The median of these absolute deviations is taken as the error $\Delta$ on the estimate of the ESS, that is:
\begin{align}
    \Delta\rho^*_D = \text{median}\left[|\rho^*_D - (\rho^*_D)_i|\right]
    \\
    \Delta\rho^*_+ = \text{median}\left[|\rho^*_+ - (\rho^*_+)_i|\right]
\end{align}

\subsection*{Unstable state (critical mass of cooperators)}
We measured numerically also $\rho^*_-$, the critical mass of cooperators needed to observe stable cooperation. Since $\rho^*_-$ is the unstable solution of the replicator equation, we need to employ a different approach than the one used for finding the  stable solution. 
First, we divide the 1000 simulation runs (see the SM section on the stable states' numerical results for details on the simulations) into 25 batches consisting of 40 runs each. For each batch $i$, we found the point $(\rho_{min})_i$ corresponding to the minimum value of the quasistationary (QS) probability distribution between the two peaks $(\rho^*_D)_i$ and $(\rho^*_+)_i$. To estimate $(\rho^*_-)_i$ we then integrate the QS probability distribution up to $(\rho_{min})_i$. The idea behind this approach is that, if a system starts in an initial condition with fewer cooperators than $\rho^*_-$, on average it will end up in (or close to) $\rho^*_D$, while if it starts with more cooperators than $\rho^*_-$, it will end up in (or close to) $\rho^*_+$. Thus, the area under the QS distribution until the minimum $(\rho_{min})_i$ is proportional to the fraction of initial conditions ending up close to $(\rho^*_D)_i$, i.e., to the size of the basin of attraction of $(\rho^*_D)_i$ which is the definition of $(\rho^*_-)_i$. 
We then computed $\rho^*_-$ as the mean value of $(\rho^*_-)_i$:
\begin{equation}
     \rho_-^* = \langle (\rho_-^*)_i \rangle
\end{equation}
where $<\cdot>$ is the average over the batches (in our case $i \in [1, 25]$). As the error on $\rho^*_-$ we instead took the standard deviation $\Delta \rho_-^* = \text{std}[(\rho_-^*)_i]$.

\section*{Details of the analytical results}
We adopt an evolutionary game theoretic approach to describe a well-mixed population of players engaged in a higher-order game. At each time step a randomly selected player (namely the focal) interacts with probability $\delta$ with other two players in a 3-person game (namely 3-game) described by the payoff tensor in Fig.~\ref{fig:payoff_tensor_colors}, while with probability $1-\delta$ it plays with another player in the pairwise version of the game (2-game). We recall that the 2-game is completely defined by the values of S and T since by definition the payoffs for mutual defection and mutual cooperation are $1$ and $0$ respectively.
The focal player can adopt the strategy (i.e. cooperation $C$ or defection $D$) of another randomly selected player, namely the model player, with a probability that is a non-decreasing function of the payoff difference between the model and focal players. 
By denoting with $\rho(t)$ the fraction of cooperators in the population at time $t$ (i.e. $1 - \rho(t)$ is the fraction of defectors), the evolution in time of the cooperators' fraction is given by the replicator equation \cite{taylor_evolutionary_1978, schuster_replicator_1983}:
\begin{equation}
    \frac{d \rho}{dt} = \rho\left[\pi_C - \langle\pi\rangle \right]\label{eq:repli_ori}
\end{equation} 
where $\langle\pi\rangle = \rho\pi_C+(1-\rho)\pi_D $ is the average payoff, and $\pi_C$ and $\pi_D$ are the expected payoffs of a cooperator and a defector respectively. Substituting the expression for $\langle\pi\rangle$ in Eq.~\ref{eq:repli_ori} we get Eq.~(1) in the manuscript, as follows: 
\begin{align}
    \frac{d\rho}{dt} =& \rho\left[\pi_C -(\rho\pi_C+(1-\rho)\pi_D)\right] \cr
    =& \rho\left[(1-\rho)\pi_C-(1-\rho)\pi_D)\right] \cr
    =& \rho(1-\rho)\left[\pi_C - \pi_D\right]\label{eq:repli_simpli}
\end{align} 
In particular, the expected payoffs for a cooperator $\pi_C$ and defector $\pi_D$ are given by:
\begin{align}
    \pi_C  =& (1-\delta)\left[ \rho + (1-\rho)S \right] + \delta \left[ \rho^2 + 2\rho(1-\rho) G + (1-\rho)^2 S \right]
    \\
    \pi_D =& (1-\delta)\left[ \rho T \right] + \delta \left[ \rho^2 T + 2\rho(1-\rho) W \right]
\end{align}
where G (\Gpayoff), W (\Wpayoff), T (\Tpayoff), and S (\Spayoff) are the elements of the payoff tensor as shown in Fig.~\ref{fig:payoff_tensor_colors}. Note that the expected payoffs are both functions of the density of cooperators $\rho$ and the fraction of 3-game interactions $\delta$. Besides the two trivial absorbing stationary states $\rho^*_D = 0$ and $\rho^*_C = 1$, Eq.~\ref{eq:repli_simpli} has two other stationary states $\rho^*_{\pm}$ for which $ \frac{d\rho}{dt} = 0$.
We introduce the quantities $a:=2(G-W)$, $b:=T-S-1$ and $c:=(a+b)$ to simplify the payoff difference as:
\begin{equation}
    \pi_C - \pi_D = -\rho^2 c \delta + \rho(c\delta - b- 2S) + S
\end{equation}
By solving $\pi_C - \pi_D = 0$ we find the non-trivial stationary solutions as,
\begin{equation}
    \rho^*_{\pm} = \frac{c \delta - b - 2S \pm \sqrt{ (c \delta - b)^2 + 4S(b+S)}}{2c\delta}\label{eq:RE_soln}
\end{equation}
It follows that when $\Delta = \left[ c \delta - b \right]^2 + 4S(b+S) \geq 0$, then $\rho^*_{\pm}$ are real valued for every $b, c,  \delta, S$. In particular, given that $\left[c \delta - b \right]^2$ is always positive, a sufficient condition for the existence of the stationary solutions is $4S(b+S)=4S(T-1)>0$, which is always satisfied for the Stag Hunt game and Chicken game.
For the Prisoner's Dilemma and the Harmony game instead $\Delta > 0$ requires that the parameters satisfy certain conditions.
If $c > 0$, these conditions are:
\begin{align}
    \delta >& \delta^{\textrm{th}}_{1}:= \frac{b+\sqrt{-4S(b+S)}}{a+b} \label{eq:delta_crit}
    \\
    \delta <& \delta^{\textrm{th}}_{2}:= \frac{b-\sqrt{-4S(b+S)}}{a+b}
\end{align}
while if $c < 0$:
\begin{align}
    \delta <& \delta^{\textrm{th}}_{1}:= \frac{b+\sqrt{-4S(b+S)}}{a+b}
    \\
    \delta >& \delta^{\textrm{th}}_{2}:= \frac{b-\sqrt{-4S(b+S)}}{a+b}
\end{align}
In particular, for the case under investigation in the manuscript, that is the Prisoner's Dilemma with $a>0$ (and hence $c>0$, see manuscript), it is easy to verify that if $d=c \delta - b - 2S <0$, then $\rho^*_{\pm} < 0$. Instead if $d=c \delta-b-2S > 0$, then $ \rho^*_{\pm} > 0$. In particular, we have $d>0$ when
\begin{equation}
    \delta>\delta^{\textrm{th}}_+ :=\frac{b+2S}{c}
\end{equation}
It can be shown that $\delta^{\textrm{th}}_2 < \delta^{\textrm{th}}_+ < \delta^{\textrm{th}}_1$ and therefore we have positive real-valued stationary solutions $0 < \rho^*_{\pm} < 1$ only for    $\delta>\delta^{\textrm{th}}_{1}$, since if $\delta<\delta^{\textrm{th}}_2<\delta^{\textrm{th}}_+$ the real-valued solutions are negative.

\section*{Alternative Bifurcation}

\begin{figure}[htp!]
    \centering
    \includegraphics[width=0.42\columnwidth]{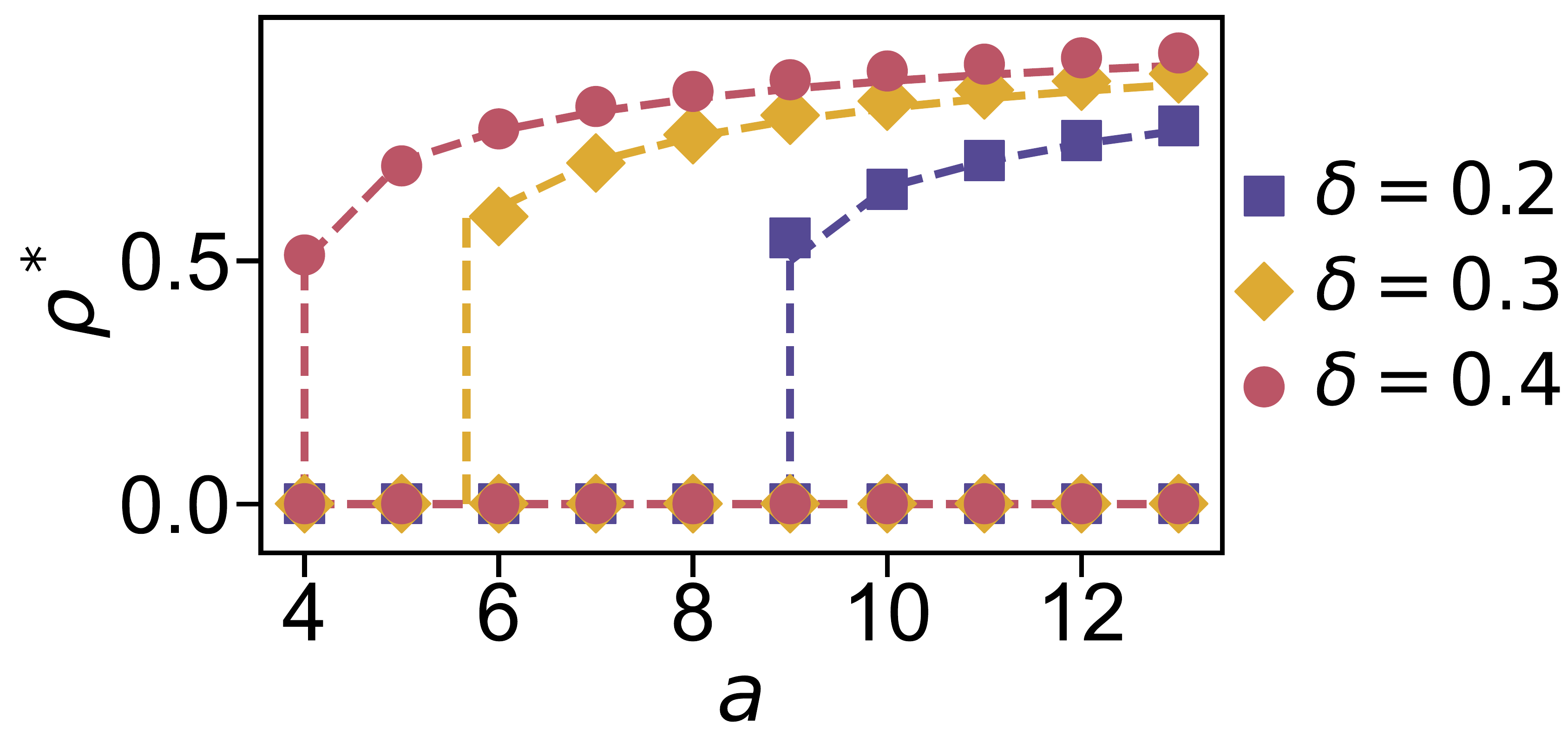}
    \caption{Fraction of cooperators at equilibrium as a function of $a$ for average hyperdegree $\langle k \rangle=20$ and different values of $\delta$. Symbols represent the numerical results averaged over 1000 independent runs (the error bars are smaller than the symbols), while dashed lines are the analytical mean-field predictions. For these results we choose $T=1.5$ and $S=-0.5$.}
    \label{fig:bifur_with_a}
\end{figure}

In the previous section of the SM, we found that the non-trivial stationary states $\rho^*_{\pm}$ described by Eq.~\ref{eq:RE_soln} exist (i.e., are real-valued and positive) iff $\delta > \delta^{\textrm{th}}_1$, where the critical threshold of 3-player interactions $\delta^{\textrm{th}}_1$ is a function of $a,b$ and $S$. 
However, the condition given by inequality Eq.~\eqref{eq:delta_crit} can also be expressed as a critical threshold on one of the other variables $a$, $b$, and $S$. For example, we can easily get a critical threshold on $a$ as a function of $\delta, b$ and $S$:
\begin{equation}
   a > a_c = \frac{b(1-\delta)+\sqrt{-4S(b+S)}}{\delta}
\end{equation}
Fig.~\ref{fig:bifur_with_a} shows the bifurcation curve as a function of $a$ for various values of $\delta$. We observe a bifurcation in the stable points of the dynamics when $a = 2(G-W)$ exceeds a critical value $a_c$. In particular, while for $a < a_c$ the only stable NE is full defection $\rho^*_D$, as in the standard pairwise PD, for $a > a_c$ we observe the emergence of a bistable behaviour where cooperation survives: besides the full defection $\rho^*_D$, a new stable state $0 < \rho^*_+ < 1$ appears due to the effect of the payoffs associated with higher-order interactions. 

\end{document}